\documentclass[conference,a4paper]{IEEEtran}
\IEEEoverridecommandlockouts
\usepackage{cite}
\usepackage{amsmath,amssymb,amsfonts}
\usepackage{newtxtext,newtxmath}
\usepackage{graphicx}
\usepackage{textcomp}
\usepackage{xcolor}
\usepackage{url}
\usepackage{array}
\def\BibTeX{{\rm B\kern-.05em{\sc i\kern-.025em b}\kern-.08em T\kern-.1667em\lower.7ex\hbox{E}\kern-.125emX}}

\begin{document}
\title{Class Weighting versus Amount Conditioning in Credit-Card Fraud Detection: A Dollar-Metric Study with a Temporal Explanation Audit}
\author{\IEEEauthorblockN{Chenyu Wu}
\IEEEauthorblockA{\textit{Duke University}\\
2080 Duke University Road, Durham, NC 27708, USA\\
wuchenyu999@outlook.com}}
\maketitle

\begin{abstract}
Credit-card fraud losses are monetary, but papers often judge models with
transaction-level scores. We ask whether transaction amount should shape training
weights or be used later to order alerts. To separate this question from ordinary class
imbalance handling, we keep total fraud-case weight fixed and vary only its allocation
across fraud cases. The experiments test two chronological card-fraud datasets with
XGBoost under unweighted training, standard class weighting, matched log-amount
weighting, stronger amount-weighted variants, and score times amount reranking. Metrics
are average precision, dollar recall, and dollar precision at fixed alert budgets over
five seeds, with 95 percent day-block bootstrap intervals for the main contrasts.
Results are narrower than expected. Amount-derived ratio and velocity features carry
much of the signal, while raw amount fields add little once those features are in the
model. In the matched setting, amount-conditioned training gives only small gains over
class weighting and does not consistently beat the plain unweighted model. Stronger
amount weights recover more fraudulent dollars, but at lower ranking quality and dollar
precision. Reranking alerts by score times amount after training gives the largest
dollar-recall shift. A small SHAP audit finds larger month-to-month attribution movement
for fraud cases than for aggregate traffic. In these tests, amount is useful as a
feature and as an alert-ordering variable, not by itself as a better sample-weighting
rule.
\end{abstract}


\section{Introduction}

Payment fraud is a costly failure mode in retail banking, and supervised learning is
standard practice for detecting it \cite{ref1,ref2}. Common model metrics rarely match
the way fraud causes loss. Precision, recall, and the area under the receiver operating
characteristic curve treat a missed five-dollar purchase and a missed
five-thousand-dollar purchase as the same event, so a model can look strong while the
dollars leaving the bank tell a different story.

One natural response is to weight fraudulent training cases by amount, so high-value
fraud has more influence during fitting. That idea is consistent with example-dependent
cost-sensitive learning \cite{ref3,ref4}, but it creates a confound. An amount-based
weight usually increases the total weight placed on the fraud class, which is also what
ordinary class weighting does. A gain credited to amount conditioning may come from a
larger positive-class weight, not from the within-fraud distribution of that weight.

This paper runs the controlled comparison. We hold the total fraud-class weight fixed
and change only its allocation across fraud cases: uniform for standard class weighting,
or proportional to log amount for amount conditioning. It is a diagnostic comparison,
not a new detector, so the contribution is the design rather than a model: a
matched-total-weight setup that separates allocation from mass, a dollar-budget
evaluation, and an explanation audit run on the alert population. Analogous concerns
arise elsewhere: separating a confounded aggregate into its components is a recurring
inference problem \cite{ref5}, and a relevant signal need not warrant the conclusion
drawn from it \cite{ref7}. Experiments report dollar recall and dollar precision at five alert
budgets, average results over random seeds, and day-block bootstrap intervals for the
main contrasts. Two reference points bound the role of amount itself: a model with
amount-derived features removed and a ranker that uses amount alone. A temporal SHAP
audit then asks whether the explanation profile shifts over time, especially within
fraud cases and high-score alerts.

Our results are deliberately qualified. Amount should remain in the modeling pipeline:
ratio and velocity features derived from amount carry signal, and dollar metrics expose
behavior that average precision can hide. Once
ordinary fraud up-weighting is controlled, matched amount conditioning adds little and
does not reliably improve on a plain model. When the objective is to recover more
fraudulent dollars under a fixed review budget, the more direct intervention is often at
inference: rank by fraud score for general detection, then use a rule such as score
times amount when dollar recall is the priority. That rule lowers precision, so it is an explicit queue policy. This SHAP audit adds a monitoring caution:
fraud-case explanations can move while the aggregate explanation plot looks stable, so
audits should include the alert population that analysts actually review.

\section{Related Work}

Credit-card fraud detection has a long data-mining literature \cite{ref1}, and much of
it is cost-sensitive. Example-dependent formulations \cite{ref4} minimize expected
financial loss instead of event-count error, cost-aware features improve fraud savings
on card data \cite{ref3}, and recent work has paired interpretable gradient boosting
with SHAP for insurance-fraud triage \cite{ref6}. Transaction amount is a recurring
input in these designs, appearing as a loss term, a sample weight, or a cost feature.
This design choice is reasonable, but it is easy to misread. If amount weighting also
raises the total positive-class weight, the reported benefit may reflect class imbalance
handling more than amount conditioning. Prior comparisons rarely hold that total fixed,
which is the gap addressed here.

Imbalance is the other long-standing difficulty \cite{ref8}, and it shapes how the
amount can be used. Resampling methods such as SMOTE \cite{ref9} and inverse-frequency
weighting are common remedies, aggregating a cardholder's recent history sharpens the
fraud signal \cite{ref3,ref10}, and sequence and hybrid models have pushed detection
accuracy further \cite{ref11,ref12}. Deep models are now competitive: graph neural
networks resist camouflage by fraudsters \cite{refDL2} or propagate risk through a gated
temporal attention graph \cite{refDL1}. We do not claim to beat these, since they need
entity graphs or sequence context beyond our tabular setting, and boosting remains the
stronger family on typical tabular data \cite{refDL3}. Plain and class-weighted gradient
boosting \cite{ref13,ref14,ref15} therefore remain strong and widely used baselines, and
they serve here as the reference points an amount-conditioned scheme has to beat.

Because fraud is rare, the evaluation metric matters as much as the model.
Precision-recall views are more informative than receiver operating characteristic views
under heavy imbalance \cite{ref16}, and the two curves are formally related
\cite{ref17}. Ranking summaries say little, though, about how much money a fixed review
budget recovers, which has motivated cost- and amount-aware evaluation \cite{ref3,ref4},
construct-aligned grading of financial models \cite{ref18}, and economic-validity audits
of tabular predictions used in discrete choice \cite{ref19,ref20}. Dollar recall and
dollar precision follow this line.

Explanation stability over time is a separate concern. SHAP \cite{ref21,ref22} and LIME
\cite{ref23} are common attribution tools; counterfactual explanations provide a
complementary action-oriented view \cite{refRELAX}, and recent fraud studies compare
several
methods on anonymized data \cite{ref24}. Concept drift is well studied in general
\cite{ref26,ref27} and in fraud \cite{ref2}. Less often checked is whether the
attribution profile itself changes across chronological periods, and whether that change
is larger in fraud cases than in ordinary traffic.

\section{Methodology}

\subsection{Datasets}

Sparkov \cite{ref28} is a synthetic credit-card dataset with interpretable fields:
amount, merchant category, cardholder and merchant coordinates, time of day, and
demographics. Using the public chronological partition, we split its training portion
again by time, giving 972,506 training transactions (5,523 fraud, 0.57 percent), 324,169
validation (1,983 fraud, 0.61 percent), and 555,719 test (2,145 fraud, 0.39 percent);
the test set holds 1.13 million dollars of fraud out of 38.6 million. The pooled rate is
about 0.52 percent, and dates run from January 2019 to December 2020. IEEE-CIS
\cite{ref29} is a
real e-commerce, card-not-present fraud dataset from Vesta spanning about six months;
most features are anonymized. A chronological 60/20/20 split gives 354,324 training
(11,988 fraud), 118,108 validation (4,611 fraud), and 118,108 test (4,064 fraud), a 3.5
percent fraud rate. IEEE-CIS is used only to test transfer of the weighting comparison, since its anonymized features make SHAP attributions hard
to read.

\subsection{Preprocessing and leakage controls}

Splits are chronological and never shuffled; the test set is evaluated once. Categorical
fields are one-hot encoded, not target encoded. Per-cardholder rolling statistics use a
one-step lag, so the current transaction cannot enter its own rolling window. For
IEEE-CIS, the row identifier and raw time offset are dropped after ordering because both
can leak position under a chronological split. High-cardinality fields are frequency
encoded from the training split only. A runtime check confirms that no identifier, raw
time field, or non-numeric column reaches the model. Weighting choices and thresholds
are selected on validation data.

\subsection{Weighting schemes}

Let $a_i$ be the transaction amount and let s =
$n_{\mathrm{negative}}/n_{\mathrm{positive}}$, computed on the training split.
Legitimate cases have weight 1 in every run. We compare four ways of weighting fraud
cases. Plain also gives fraud cases weight 1. ClassWt uses $w_i$ = s. AmtLogN uses $w_i$
= s $\log(1+a_i)$ / $\mathrm{mean}_{\mathrm{train,fraud}}[\log(1+a)]$. Because its mean
fraud weight is s, ClassWt and AmtLogN assign the same total fraud weight; only the
allocation among fraud cases changes. That is the comparison we care about most. AmtLog
uses $w_i$ = 10 $\log(1+a_i)$, so its total fraud weight is not matched to ClassWt.
AmtLin uses $w_i$ = $a_i$ only as a stress test.
We test the log shape: amounts are heavy-tailed, so a linear
weight lets a few large frauds dominate the gradient while the log compresses that tail.
To check that the compression rather than the specific function does the work, we add
matched AmtSqrtN and AmtRankN variants replacing $\log(1+a)$ by $\sqrt{a}$ and by the
within-fraud rank of $a$.

Two reference runs help keep the interpretation honest. NoAmt removes amount-derived
features. AmtOnly ranks by amount alone. On IEEE-CIS, NoAmt removes only TransactionAmt
because anonymized variables may still encode amount-related information. Logistic
regression and random forest baselines with balanced class weights \cite{ref15} reached
Sparkov average precision 0.22 and 0.89, below the boosting models. Sample weights can
also change effective regularization, so the matched setup and the depth sweep check,
roughly, that we are not calling a regularization effect an amount effect. At inference
we also rerank Plain by multiplying its score by raw dollar amount. Because XGBoost
scores are not calibrated probabilities, we use the product only as a ranking signal.

\subsection{Evaluation protocol}

Average precision is computed with scikit-learn's average\_precision\_score summary of
the precision-recall curve \cite{ref16,ref17}. For queue metrics, the alert budget is the
top k percent of highest-scoring transactions. We evaluate k = 0.1, 0.5, 1, 2, and 5
percent, and report representative budgets in the tables. Dollar recall is the share of
fraudulent dollars captured. Dollar precision is the share of flagged dollars that are
fraudulent. Recorded outputs also include the false-positive count and the mean
legitimate amount flagged. Each model is trained under five random seeds, and tables
report the mean and standard deviation. For the key contrasts, a block bootstrap
resamples whole calendar days because same-day transactions are correlated; we draw
1,000 resamples conditional on the seed-42 models and report 95 percent intervals.
Because several are inspected at once, interval widths give normal-approximation
standard errors and two-sided p-values, corrected with Holm--Bonferroni across the
twelve contrasts. Contrast estimates and intervals use the seed-42 models while tables
report five-seed means, so the treatment remains descriptive.

\subsection{Temporal SHAP drift protocol}

Our audit uses the log-amount model (AmtLog). SHAP values are computed with the
tree-path-dependent estimator \cite{ref22}, which uses the tree structure and needs no
background dataset. For each calendar month, importance is the mean absolute SHAP value
over up to 2,000 sampled transactions, or all available cases when a subset is smaller.
That exception matters for fraud months, where counts are often low. Importance vectors
are normalized to sum to one before comparison. A fixed-model design trains once on the
first three months and explains each later month, measuring data-driven change; three
months is a pragmatic minimum, leaving most of the timeline for auditing. An
expanding-window design retrains each month before explaining, adding model-update
change. Across the 21 audited months we measure Jensen-Shannon distance between
consecutive months and Kendall rank correlation against the first month, for all
transactions, true fraud cases, and the highest-scoring one percent.

\section{Experiments And Results}

\subsection{Setup}

All gradient boosting models use 500 trees, depth seven, learning rate 0.05, and 0.8
column and row subsampling, trained with the histogram method \cite{ref13}. Experiments
run on a laptop with 16 GB of memory and no graphics accelerator. On validation, we set
the AmtLog multiplier to ten. A depth sweep over 5, 7, and 9 on the Sparkov validation
set keeps the schemes within 0.006 average precision of one another at every depth, so
the depth-seven choice does not favor one weighting scheme. Seeds are fixed, library
versions are pinned (XGBoost 2.1.4, SHAP 0.49.1), splits are deterministic and
chronological, and exact split sizes are reported. Cost is set by the data and the tree
budget, not by the weighting: median fit time is 69 s on Sparkov and 50 s on IEEE-CIS,
and scoring runs at 3.3 and 4.5 microseconds per transaction. Reranking adds no
measurable cost; the SHAP audit is the expensive part and runs as a scheduled monthly
job, and tighter budgets may draw on analogous work in compression and split execution
\cite{ref30} or resource-aware scheduling \cite{ref34}. Deployments also carry
review-capacity and privacy constraints we omit; analogous online dispatch work models
capacity and privacy jointly \cite{ref37}.
Code and configuration are available from the authors and will be released in a public
repository on acceptance.

\subsection{Amount as a feature}

Removing all amount-derived features collapses ranking on Sparkov, from 0.930 average
precision to 0.336, but the raw amount column alone is not the cause: removing it leaves
0.907, and removing only the velocity features, the seven-day rolling mean and the
amount-to-average ratio, leaves 0.874. Ranking collapses only when both go. Explicit
amount columns are largely redundant with the derived ones, since the ratio already
contains the current amount. On IEEE-CIS, removing the amount field changes average
precision only from 0.548 to 0.538, although anonymized features may still carry
amount-related signal. Ranking by amount alone is weak on both datasets (0.14 on
Sparkov, 0.04 on IEEE-CIS), so trained models are not simply sorting by size. Amount is
useful as a feature; that is separate from whether it should set sample weights.

\subsection{Controlled weighting comparison}

Tables~\ref{tab:sparkov} and~\ref{tab:ieee} give the five-seed means, with standard
deviations for AP and DR@1\%. On Sparkov, the boosting schemes are close in both average
precision and dollar recall. AmtLogN is 0.002 above ClassWt in average precision
(day-block interval 0.0002 to 0.004). That difference is nominally detectable but does
not survive the Holm correction over the twelve contrasts (adjusted p = 0.20), and it is
too small to matter in an alert queue. For DR@1\%, the interval includes zero. Against
Plain, neither interval excludes zero. Sparkov is highly separable at small alert
budgets, so there is little room for any weighting scheme to improve detection. A more
visible movement is DP@1\%, which rises from 0.47 for Plain to 0.58 for ClassWt, with
AmtLogN also at 0.58: the reduction in false-positive dollars comes from fraud-class
weighting, not from conditioning that weight on amount. At the 1\% budget this is about
3,500 false positives and a mean legitimate amount flagged of 235 dollars for ClassWt
and AmtLogN, against 368 dollars for Plain.

\begin{table}[!tb]
\caption{Sparkov test results, mean over five seeds (std in parentheses for AP and DR@1\%). AP: average precision; DR: dollar recall; DP@1\%: dollar precision at 1\%.}
\label{tab:sparkov}
\centering
\scriptsize
\setlength{\tabcolsep}{2pt}
\resizebox{\columnwidth}{!}{%
\begin{tabular}{lccccc}
\hline
\textbf{Model} & \textbf{AP} & \textbf{DR@0.5\%} & \textbf{DR@1\%} & \textbf{DR@2\%} & \textbf{DP@1\%} \\
\hline
Plain & 0.930 (.001) & 0.970 & 0.990 (.001) & 0.997 & 0.47 \\
ClassWt & 0.931 (.001) & 0.973 & 0.992 (.000) & 0.997 & 0.58 \\
AmtLogN (matched) & 0.932 (.000) & 0.973 & 0.992 (.001) & 0.997 & 0.58 \\
AmtLog & 0.934 (.000) & 0.974 & 0.993 (.001) & 0.997 & 0.57 \\
AmtLin & 0.927 (.001) & 0.978 & 0.991 (.002) & 0.997 & 0.58 \\
Plain + score*amt rerank & 0.820 & 0.980 & 0.995 & 0.998 & 0.25 \\
\hline
\end{tabular}%
}
\par\vspace{1mm}
\begin{minipage}{\columnwidth}
\footnotesize NoAmt and AmtOnly have DR@1\% of 0.63 and 0.83; at the 0.1\% budget,
precision is one.
\end{minipage}
\end{table}

IEEE-CIS gives a different picture. Plain is the strongest baseline among the tested
models, with average precision 0.548 and DR@1\% of 0.188. None of the amount-conditioned
schemes beats that dollar recall with an interval excluding zero. ClassWt is harmful
here: average precision falls to 0.513 and DR@1\% to 0.155. AmtLogN recovers part of
that loss. Its DR@1\% gain over ClassWt is 0.020 (day-block interval 0.007 to 0.034),
while the average-precision interval includes zero (interval -0.005 to 0.007). Against
Plain, though, AmtLogN is still lower on average precision (interval -0.042 to -0.025),
and its DR@1\% interval includes zero. That DR@1\% gain is the one matched result
surviving Holm correction (adjusted p = 0.032); six of seven nominally significant
contrasts survive. Because model specification can reorder rankings, as cross-market
volatility forecasting illustrates \cite{ref36}, we
repeated the comparison in a second boosting implementation, LightGBM, which shows the
same pattern: average precision 0.555 for Plain against 0.515 for ClassWt and 0.514 for
AmtLogN, the latter two differing by 0.001, and AmtLogN again above ClassWt on dollar
recall (0.175 against 0.158).

\begin{figure}[!tb]
\centering
\includegraphics[width=\columnwidth]{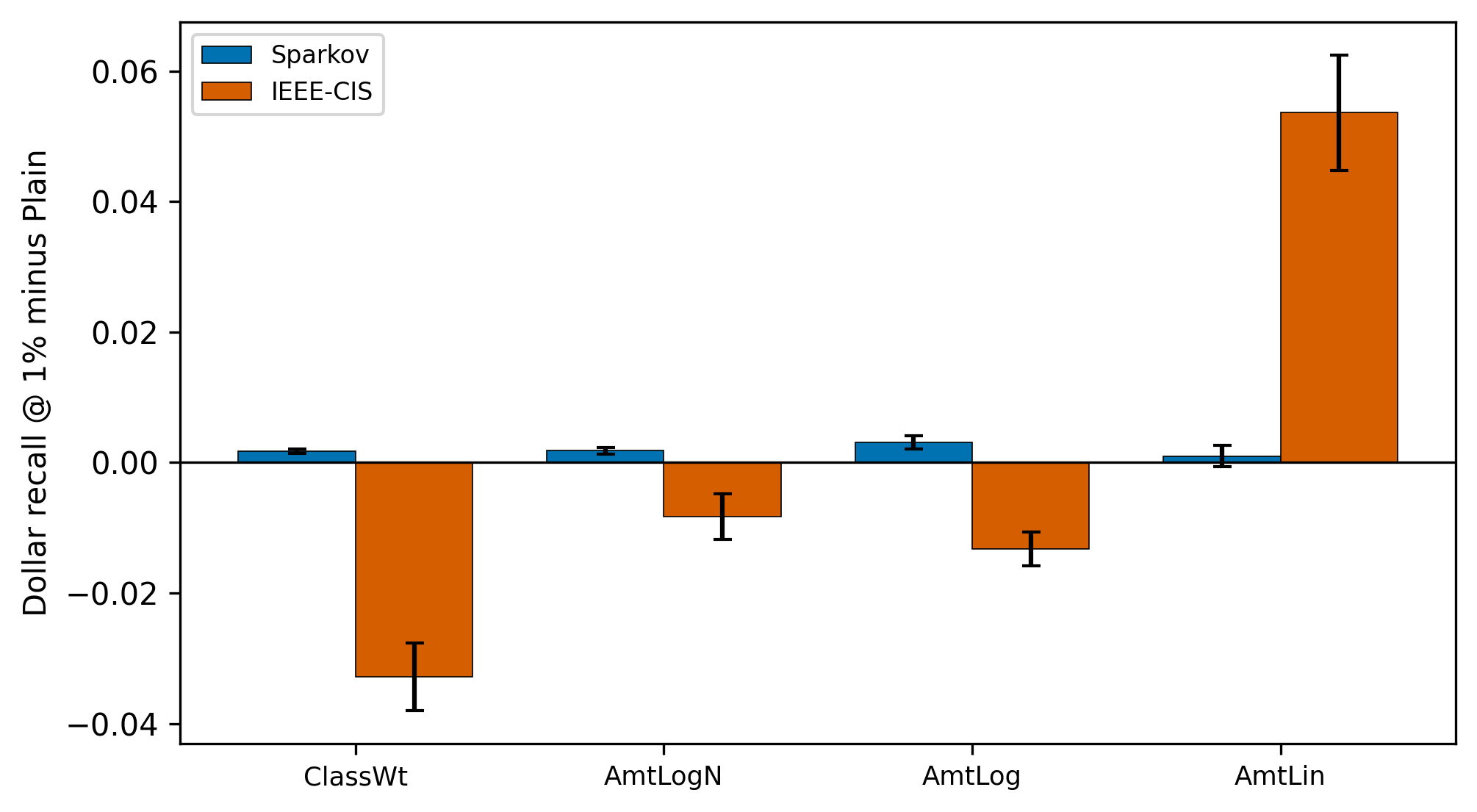}
\caption{Dollar recall at the 1\% budget for each scheme minus the unweighted model, mean over five seeds with std bars.}
\label{fig:delta}
\end{figure}

Among trained schemes, only AmtLin has a dollar-recall increase over Plain whose
interval excludes zero. It raises dollar recall by 0.054 (interval 0.033 to 0.076), but
average precision drops by 0.108 (interval -0.124 to -0.091) and DP@1\% falls from 0.89
to 0.68. A simpler reranking check is more revealing. Multiplying the Plain score by
amount raises DR@1\% to 0.384, above every trained scheme, while average precision falls
to 0.31 and DP@1\% to 0.33. We read this as a different alert ordering. Tempering the exponent traces that trade-off: with score times
amount$^{\gamma}$ at $\gamma$ = 0.25, 0.5, and 0.75, average precision is 0.520, 0.457,
and 0.380 at DR@1\% of 0.279, 0.335, and 0.366, so $\gamma$ = 0.5 keeps most of the
dollar-recall gain at far better ranking, exceeding every trained scheme on DR@1\%
though at lower average precision than all but AmtLin. A flat per-review cost sets the
budget, not the order: thresholding the product at a constant leaves the ordering
unchanged. Calling it cost-optimal would require calibrated probabilities, which we do
not claim. Reranking as a post-hoc stage is familiar
elsewhere \cite{ref33}, and the expected-value view resembles portfolio rules trading
return against downside risk \cite{ref31}. Fig.~\ref{fig:delta} reports dollar recall
relative to Plain.

\begin{table}[!tb]
\caption{IEEE-CIS test results, mean over five seeds (std in parentheses for AP and DR@1\%). Columns as in Table~\ref{tab:sparkov}.}
\label{tab:ieee}
\centering
\scriptsize
\setlength{\tabcolsep}{2pt}
\resizebox{\columnwidth}{!}{%
\begin{tabular}{lccccc}
\hline
\textbf{Model} & \textbf{AP} & \textbf{DR@0.5\%} & \textbf{DR@1\%} & \textbf{DR@2\%} & \textbf{DP@1\%} \\
\hline
Plain & 0.548 (.002) & 0.092 & 0.188 (.004) & 0.344 & 0.89 \\
ClassWt & 0.513 (.002) & 0.076 & 0.155 (.005) & 0.310 & 0.89 \\
AmtLogN (matched) & 0.514 (.003) & 0.082 & 0.180 (.004) & 0.345 & 0.89 \\
AmtLog & 0.503 (.004) & 0.080 & 0.175 (.003) & 0.340 & 0.89 \\
AmtLin & 0.444 (.004) & 0.131 & 0.242 (.009) & 0.379 & 0.68 \\
Plain + score*amt rerank & 0.307 & 0.274 & 0.384 & 0.516 & 0.33 \\
\hline
\end{tabular}%
}
\par\vspace{1mm}
\begin{minipage}{\columnwidth}
\footnotesize Rerank multiplies Plain scores by amount without retraining.
\end{minipage}
\end{table}

\subsection{Sensitivity to weighting strength}

\begin{figure}[!tb]
\centering
\includegraphics[width=\columnwidth]{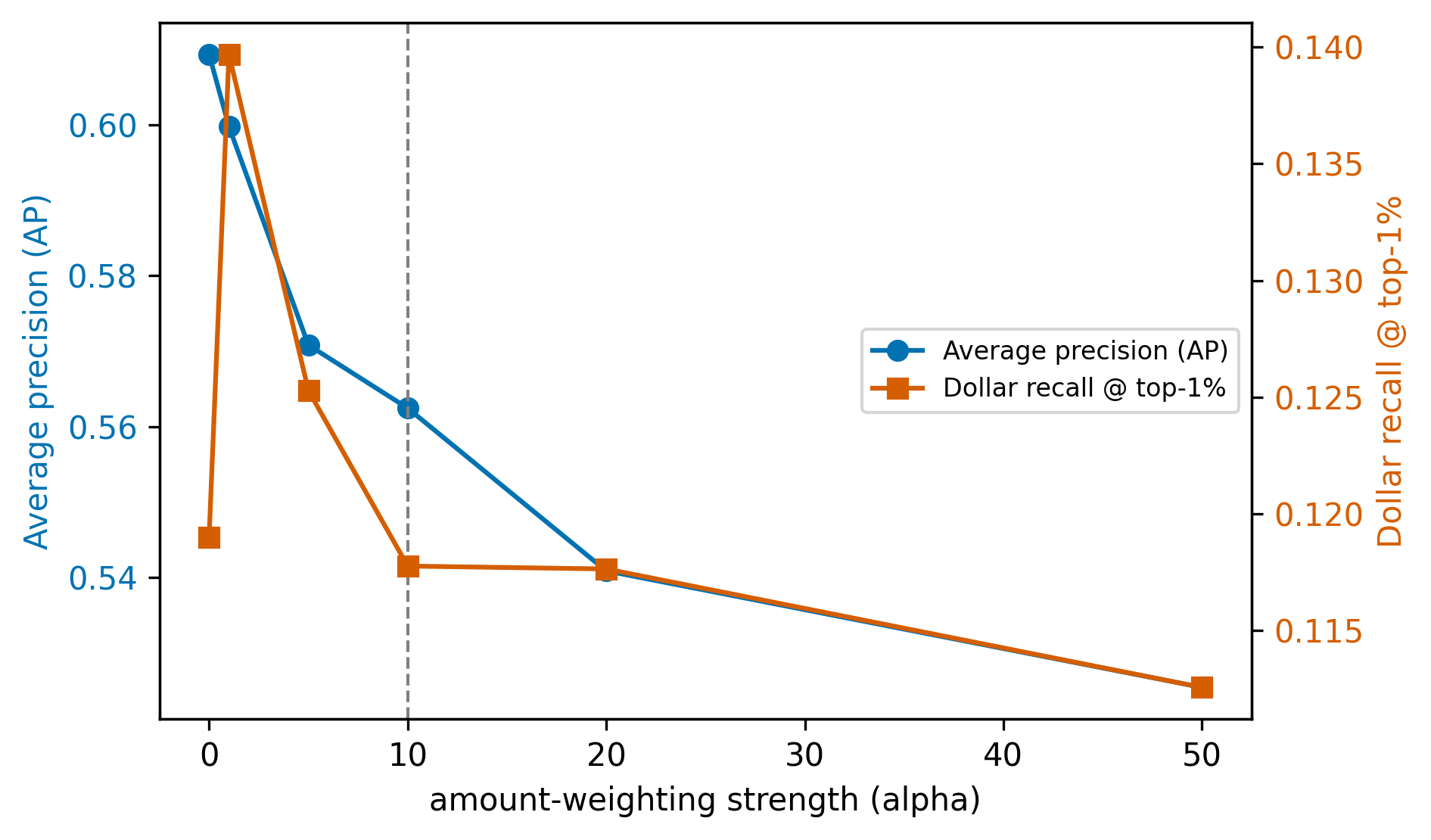}
\caption{IEEE-CIS validation performance as the fixed log-amount multiplier varies (the AmtLog family, distinct from the matched-weight AmtLogN).}
\label{fig:sensitivity}
\end{figure}

To map the weighting-strength trade-off, we swept the fixed multiplier of the log-amount
weight on the IEEE-CIS validation set. That sweep is the AmtLog family. Average precision falls steadily as the multiplier grows,
from 0.609 with no weighting to 0.525 at the largest setting. DR@1\% peaks at multiplier
one, so the knob is not monotone in the metric it is meant to improve; non-monotone
returns from a single scaling parameter are also reported in LLM tuning
\cite{ref32,ref35}. The limited checks suggest shape matters less than compression:
matched square-root and rank
variants both reach 0.930 average precision on Sparkov against 0.932 for AmtLogN, and on
IEEE-CIS the schemes trace one axis ordered by tail compression (ClassWt 0.513 at DR@1\%
0.155, AmtLogN 0.514 at 0.180, AmtSqrtN 0.504 at 0.214, AmtLin 0.444 at 0.242), none
reaching Plain's 0.548. AmtLogN is used for the controlled comparison: AmtLog changes
both the within-fraud allocation and the total fraud weight, while AmtLogN holds the
total fixed.
Fig.~\ref{fig:sensitivity} shows the validation trend.

\subsection{Temporal explanation drift}

\begin{figure}[!tb]
\centering
\includegraphics[width=\columnwidth]{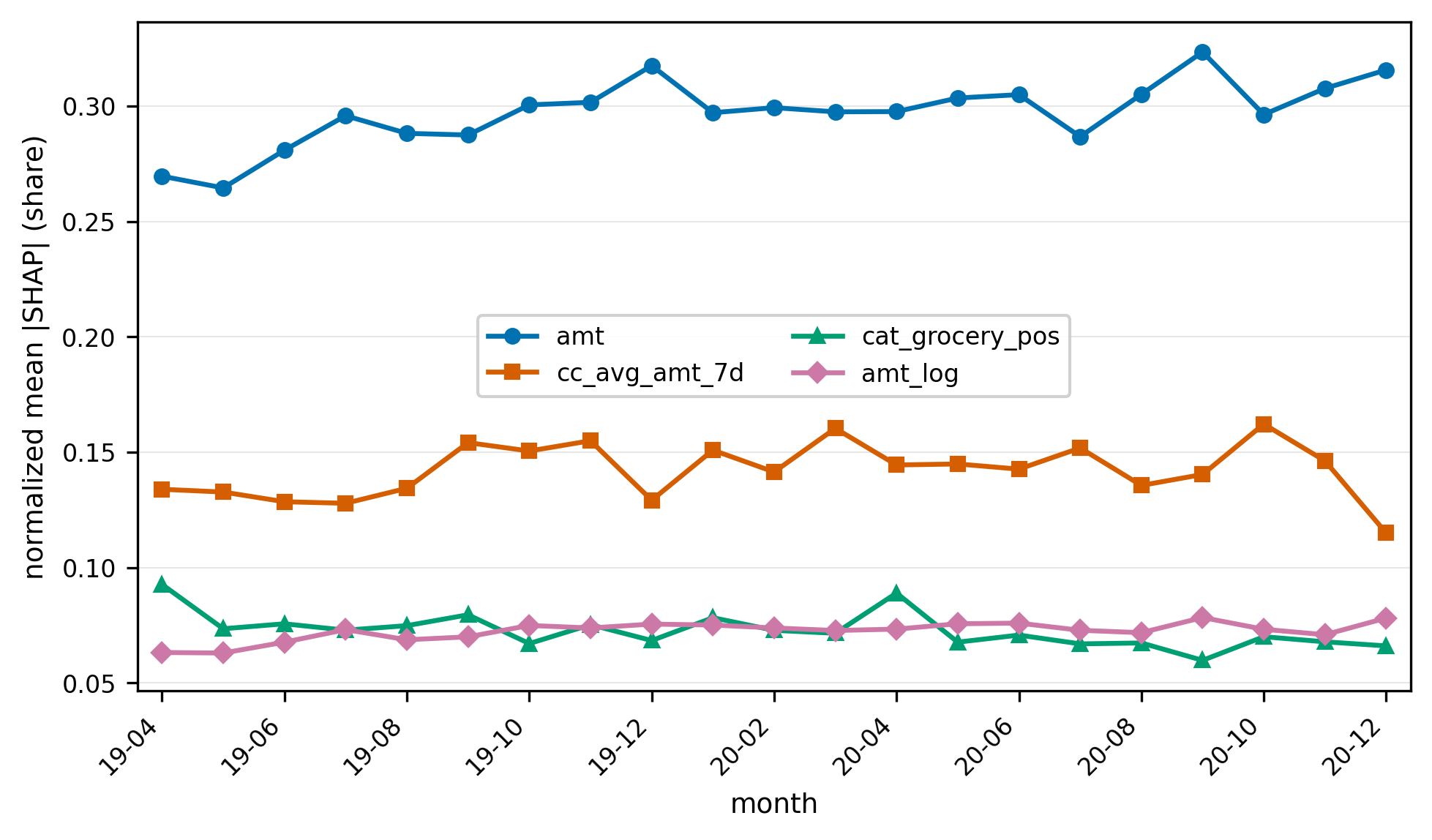}
\caption{Normalized monthly SHAP importance of the leading fraud-subset features (expanding-window design, Sparkov, log-amount model).}
\label{fig:drift}
\end{figure}

Table~\ref{tab:drift} reports the Sparkov drift audit for AmtLog, one representative
amount-aware classifier. We did not repeat the audit for every scheme; this part of the
paper is exploratory. Score times amount only reorders Plain scores, so it has no
separate trained model to explain. Fraud-subset explanations move much more than
all-traffic explanations, with mean Jensen-Shannon distance 0.057 compared with 0.012 in
the fixed-model design; an aggregate curve can stay flat while local behavior moves
underneath it, as also reported for model editing \cite{ref25}. The 2,000-transaction cap does not drive this: monthly fraud
counts run from 258 to 592, so it never binds on the fraud subset, and recomputing the
all-traffic curve at caps of 1,000, 2,000, and 5,000 gives 0.017, 0.012, and 0.009,
leaving fraud drift over three times the aggregate at every cap. In this run, expanding-window retraining increases measured drift on every subset. Fraud-subset monitoring is retrospective
because labels arrive with delay; the top-one-percent alert subset is easier to compute
immediately, although it changes when the model changes and so mixes explanation drift
with selection drift. The audit matters because dollar-oriented operation makes
fraud-case monitoring relevant, and fraud-case explanations moved most.
Fig.~\ref{fig:drift} plots normalized monthly importance.

\begin{table}[!tb]
\caption{Temporal SHAP drift on Sparkov (log-amount model). Higher JS distance and lower Kendall tau mean more drift.}
\label{tab:drift}
\centering
\scriptsize
\setlength{\tabcolsep}{2pt}
{%
\begin{tabular}{lccc}
\hline
\textbf{Design} & \textbf{Subset} & \textbf{Mean JS} & \textbf{Min Kendall tau} \\
\hline
A (data only) & all & 0.012 & 0.97 \\
A (data only) & fraud & 0.057 & 0.82 \\
A (data only) & top 1\% & 0.043 & 0.89 \\
B (retraining) & all & 0.037 & 0.71 \\
B (retraining) & fraud & 0.074 & 0.68 \\
B (retraining) & top 1\% & 0.061 & 0.72 \\
\hline
\end{tabular}%
}
\par\vspace{1mm}
\begin{minipage}{\columnwidth}
\footnotesize A uses a fixed model trained on months 1--3; B uses monthly
expanding-window retraining.
\end{minipage}
\end{table}

\section{Discussion}

Our matched-weight results do not support amount weighting as the default move. Amount
is not irrelevant: ratio and velocity variables carry much of the useful signal, and
dollar metrics expose behavior average precision misses. On Sparkov the small-budget task is nearly solved, so the schemes have little room
to separate. On IEEE-CIS, Plain is the best baseline; ClassWt hurts it and AmtLogN
repairs some of that damage without moving ahead. AmtLin buys more dollar recall, yet a
post-training reranking rule moves farther along the same trade-off without retraining.
For deployment the first decision is queue design: train a strong baseline, then choose
an alert ordering matching the review goal. Amount-conditioned training is still worth
testing where class weighting or an explicit cost model is already in use.

The SHAP audit gives a smaller warning. Dollar-aware rules change the alert population;
in the audit, fraud-case explanations moved fastest, and a single global SHAP plot would
have missed it.
Timing remains a problem: confirmed labels arrive late, so the fraud subset is
retrospective, while top-one-percent subsets are immediate but shift when the model
updates.

\section{Limitations}

Scope here is limited. Sparkov is synthetic; its named variables suit controlled checks,
but the near-perfect small-budget precision should not be read as live alert quality.
IEEE-CIS is a useful transfer check, although its anonymized fields make feature-level
interpretation weak, and the relative ordering of schemes is more informative than
absolute scores. We study only gradient-boosted trees, in two libraries; a genuinely
different model family, sequence or graph models, or richer entity features may change
the comparison. These weighting rules are also simpler than deployed cost models,
omitting review cost, recovery, customer friction, and intervention effects. Sample
weights affect regularization, which the matched design and depth checks address only
partly. Our statistical treatment is
descriptive: corrected p-values are normal approximations from percentile intervals
conditional on one seed, and SHAP drift uses monthly point estimates without confidence
bands. Delayed labels, alert-selection
effects, and rules for refreshing explanation baselines are left unresolved.

\section{Ethical And Data Considerations}

Both datasets are public and contain no personally identifying information we access:
Sparkov is synthetic and IEEE-CIS anonymized. Deployed fraud models can produce unequal
error rates across customer groups and can inconvenience legitimate customers through
false alarms; our dollar-precision metric speaks to the second concern but not the
first, and any deployment should include fairness review and human adjudication. For graph-based deployments,
unlearning requests can also degrade group fairness \cite{refFROG}.

\section{Conclusion}

We asked a deliberately small question: after total fraud-case weight is fixed, does
distributing that weight by transaction amount help? In the matched setting tested here,
usually not. Amount still matters, but more through features and queue ordering than
through matched sample weights: on Sparkov, ratio and velocity features carried much of
the signal,
and score times amount reranking changed dollar recall more directly than retraining
did, while matched amount-weighted training gave only small gains over class weighting.
The caution matters, since that reranking is not a free improvement: it raised dollar
recall on IEEE-CIS but lowered average precision and dollar precision, so it is a
queue-management rule rather than a better classifier.

For papers in this area, dollar-level queue metrics belong beside transaction-level
ranking metrics, because the two views can rank methods differently, and formal claims
over a family of contrasts should carry a multiplicity correction: the Sparkov
AmtLogN-ClassWt gap in average precision does not survive one. Amount-conditioned training may
still help where class weighting or explicit cost models are already in use, but
comparisons should include Plain and inference-time reranking. Richer entity features
and explicit investigation costs would make the cost setting more realistic, and for the
explanation audit delayed labels and confidence bands are the next missing pieces.


\end{document}